-------------------------------------------------------------------------

\documentstyle[eqsecnum,aps,epsf]{revtex}      

\begin{document}
\twocolumn[\hsize\textwidth\columnwidth\hsize\csname @twocolumnfalse\endcsname

\title{\bf 
Comment on ``Time-reversal symmetry breaking superconductivity"}

\author{Sadhan K. Adhikari}
\address{Instituto de F\'{\i}sica Te\'orica, 
Universidade Estadual Paulista,
01.405-900 S\~ao Paulo, SP, Brazil \thanks{permanent address}\\
Instituto de F\'{i}sica,  Universidad Nacional Aut\'{o}noma de
M\'{e}xico,
  01.000 M\'{e}xico City,  DF,  M\'{e}xico}

\date{\today}

\maketitle

\begin{abstract}

It is pointed out that erroneous Bardeen-Cooper-Schrieffer model equations have been
used by Haranath Ghosh in his recent treatment of time-reversal symmetry-breaking
superconductivity. Consequently, his numerical results are misleading, and his
conclusions are not to the point.

{PACS number(s): 74.20.Fg, 74.62.-c, 74.25.Bt}

\end{abstract} 

\vskip1.5pc]

\newpage

Some recent studies provide increasing
evidence that the pairing symmetry of some of the cuprates at low 
temperatures allows an order parameter in a mixed symmetry state.
At higher temperatures, below the critical temperature $T_c$,  the
symmetry of the order parameter is of the $d_{x^2-y^2}$ type. At a lower
temperature there could be an admixture of a minor component, such as 
 $d_{xy}$, on the predominant $d_{x^2-y^2}$ symmetry. This general 
time-reversal symmetry breaking  order parameter has the form 
$d_{x^2-y^2} + \exp(i\theta)d_{xy}$, where $
\theta$ is the mixing angle. 

Recently, Ghosh \cite{gh1} presented a theoretical study of
superconductivity for this mixed-symmetry case based on the
Bardeen-Cooper-Schrieffer (BCS) equation.
In this comment we point out  that the coupled  equations used by him  for
the two
components of the order parameter    are erroneous and   present  a
rederivation of the appropriate equations. This comment also  applies to 
 latter investigations  where 
Ghosh
further
used the erroneous equations in    subsequent studies of (1) 
superconductors
of 
mixed order parameter symmetry in a Zeeman magnetic field \cite{gh2} and 
(2) pairing symmetry and long-range pair potential in a weak-coupling
theory of superconductivity \cite{gh3}.

We  use the 
 two-dimensional  tight binding model as in Ref. \cite{gh1}. 
The effective interaction
after including the two appropriate basis functions is
taken as 
\begin{equation}\label{2}
V_{{\bf k}{\bf q}}=-V_1\eta_{1{\bf k}} \eta_{1{\bf q}}- V_2 \eta_{2{\bf
k}}\eta_{2{\bf q}},
\end{equation}
where $\eta _{1\bf q} = \cos q_x -\cos q_y $ corresponds to $d_{x^2-y^2}$
symmetry, $\eta _{2\bf q}= \sin q_x \sin q_y $ corresponds to $d_{xy}$
symmetry.
The orthogonal functions $\eta_{1{\bf q}}$ and $\eta_{2{\bf q}}$
 are associated with a one-dimensional irreducible representation of the
point group of square lattice $C_{4v}$ \cite{xx} and can be considered
appropriate
generalizations
of the circular harmonics $\cos (2\phi)$ and $\sin (2\phi)$
incorporating the proper lattice symmetry. The orthogonality condition
of these functions is
\begin{equation}\label{ix}
\sum_{\bf q}{ \eta_{1{\bf q}}\eta_{2{\bf q}}}=0.
  \end{equation}
This orthogonality relation is readily verified as under the 
transformation
$q_x \to -q_x$  (or   $q_y \to -q_y$) $ \eta_{2{\bf q}}$ changes
sign
and $ \eta_{1{\bf q}}$ remains unchanged. Although, the proof could be
slightly different, similar orthogonality relation exists between
the basis
functions of states on square lattice, such as, $d_{x^2-y^2}$, $s$ $(\eta 
=1)$,
$s_{x^2+y^2}$ $ ( \eta _{\bf q} = \cos q_x +\cos q_y )$, and
$s_{xy}$ $( \eta _{\bf q} = \cos q_x \cos q_y ) $, and the
present discussion equally applies to mixtures involving such
orthogonal states. On the continuum the $s$-wave angular function  
$\zeta_1(\phi) =1$, and the $d$-wave circular harmonics $\zeta_2(\phi) =
\cos
(2\phi)$
and $\zeta_3(\phi)=\sin (2\phi)$ satisfy the trivial orthogonality
relation
\begin{equation}
\int_0^{2\pi} \zeta_i(\phi) \zeta_j(\phi) d\phi=0, \quad i\ne j.
\label{con} \end{equation}
One passes from the lattice to continuum description by replacing (a) the
sum
over ${\bf q}$ by  an integral over $\phi$ and (b) the functions
$\eta_{\bf q}$ by the circular harmonics $\zeta (\phi)$.

Although Ghosh \cite{gh1} considered the BCS model at a 
 a finite temperature, we consider its zero temperature version, 
which is enough for our purpose:
\begin{eqnarray}
\Delta_{\bf k}& =& -\sum_{\bf q} V_{\bf kq}\frac{\Delta_{ \bf q}}{2E_{\bf
q}},
  \label{130} \end{eqnarray} with $E_{\bf q} =
[(\epsilon_{\bf q} - \mu )^2 + |\Delta_{\bf q}|^{2}]^ {1/2},$ where 
$\epsilon_{\bf q}$ is the single-particle energy and   $\mu$ is
the chemical potential.  The order
parameter has  the following general anisotropic form: 
\begin{equation}
\Delta _{\bf q} \equiv \Delta_1 \eta _{1\bf q}
 +C\Delta_2 \eta _{2\bf q}, \label{op}\end{equation} 
where $C\equiv \exp(i\theta) = (a+ib) $ is a complex number of unit
modulus $|C|^ 2=1 $
and $a \equiv \cos\theta$ and $b\equiv \sin\theta$ are real numbers. If we
substitute 
Eqs. (\ref{2}) and (\ref{op}) into the BCS equation (\ref{130}),
for orthogonal functions $\eta_{1{\bf q}}$ and $\eta_{2{\bf q}}$,
one can separate the resultant equation into  the following components 
\begin{eqnarray}\label{a}
\Delta_1&=&V_1\sum_{\bf q}\frac{ \eta_{1{\bf q}}[\Delta_1 
\eta_{1{\bf q}}+(a+ib)\Delta_2 \eta_{2{\bf q}}
]}{2E_{\bf q}}, \\ \label{b}
(a+ib)\Delta_2&=&V_2\sum_{\bf q} \frac{\eta_{2{\bf q}}[\Delta_1 
\eta_{1{\bf q}}+(a+ib)\Delta_2 \eta_{2{\bf q}}
]}{2E_{\bf q}}.
\end{eqnarray} 
Equations (\ref{a}) and (\ref{b})  have solution for real $\Delta_1$
and
$\Delta_2$, when the complex parameter $C$ is either purely real or purely 
imaginary.  

Equations (\ref{a}) and (\ref{b}) can be substantially
simplified 
 for a purely imaginary $C$, e.g., for $C=i$ or $a=0$ and
$b=1$ ($\theta=\pi/2$).
In this case for real
components $\Delta_1$ and $\Delta_2$, the real and imaginary parts of Eqs.
(\ref{a}) and 
(\ref{b}) become, respectively, 
\begin{eqnarray}\label{a1}
\Delta_1&=&V_1\sum_{\bf q}\frac{ \eta_{1{\bf q}}^2\Delta_1 
}{2E_{\bf q}}, \\ \label{b1}
\Delta_2&=&V_2\sum_{\bf q} \frac{\eta_{2{\bf q}}^2\Delta_2 
}{2E_{\bf q}}.
\end{eqnarray}
Here we have used the identity
\begin{equation}\label{i}
\sum_{\bf q}\frac{ \eta_{1{\bf q}}\eta_{2{\bf q}}}{2E_{\bf q
}}=0,  
  \end{equation}
which holds  in this case as $ E_q \equiv [(\epsilon _q -\mu)^2+
\Delta_1^2 \eta_{1{\bf q}}^2+ \Delta_2^2 \eta_{2{\bf q}}^2]^{1/2}$ 
is 
invariant under transformation $q_x \to -q_x$ or under $q_y \to -q_y$,
whereas 
under either of these  transformations $ \eta_{2{\bf q}}$ changes
sign
and $ \eta_{1{\bf q}}$ remains unchanged. Hence
using
the  integration of ${\bf q}$, one can establish identity
   (\ref{i}).
 Equations (\ref{a1}) and (\ref{b1}) have been  used in the
study of the mixed-symmetry states of  types $d_{x^2-y^2}+id_{xy}$
and $d_{x^2-y^2}+is$
\cite{ad1}.

Equations (\ref{a}) and (\ref{b}) also lead to a simple form for a purely
real $C$,
 e.g., for $C=1$ or for $a=1$ and $b=0$ ($\theta =0$). 
However, in this case 
$ E_q \equiv [(\epsilon _q -\mu)^2+(
\Delta_1 \eta_{1{\bf q}}+ \Delta_2 \eta_{2{\bf q}})^2]^{1/2}$
contains cross terms of the type $\eta_{1{\bf q}}\eta_{2{\bf q}}$,
and is not 
invariant under transformation $q_x \to -q_x$ or under $q_y \to -q_y$.
 Consequently, Eq. (\ref{i}) is
not satisfied and coupled angular terms will be present in the BCS
equation.
In this case for real $\Delta_1 $  and $\Delta _2$,
Eqs. (\ref{a}) and (\ref{b}) become 
 the following  set of coupled equations
\begin{eqnarray}
\Delta_1=V_1\sum_{\bf q}\frac{\eta
 _{1\bf q}[\Delta_1\eta _{1\bf q}+\Delta_2
\eta _{2\bf q}]}{2E_{\bf
q}},\label{a2}
\end{eqnarray}
\begin{eqnarray}
\Delta_2=V_2\sum_{\bf q}\frac{\eta _{2\bf q}[\Delta_1\eta _{1\bf q}
+\Delta_2
\eta _{2\bf q}]}{2E_{\bf
q}}.\label{b2}
\end{eqnarray}

In the  case of a general mixture, e.g., $a\ne 0$ and $b\ne 0$,  Eq.
(\ref{i}) is not valid. However, for real  $\Delta_1$ and 
$\Delta_2$, 
one can break up Eqs. (\ref{a})  and (\ref{b}) into their real and
imaginary parts, e.g., into four coupled equations for two unknowns 
$\Delta_1$ and $\Delta_2$. As Eq. (\ref{i}) does not hold in this general
case, the four coupled equations are consistent only if 
$\Delta_2 =0$, or 
$\Delta_1 =0$, 
which corresponds to no coupling between the two
components. Hence the permissible values for the mixing angle $\theta$
are 0, $ \pi/2$, $\pi$, and $3\pi/2$.

In his study Ghosh \cite{gh1} implicitly assumed Eq. (\ref{i}) to be valid
in all
the cases discussed above including (i) the case of a general $C $ with
$a\ne
0$ and $b\ne 0$, and (ii) the case with $a=1$ and $b=0$. Consequently, he 
 arrived at the wrong equations (\ref{a1}) and (\ref{b1}) for a general
$C$,
 which he used in his numerical treatment, specifically for mixing angles
$\theta$ = 0, and $\pi/4$ (see Eq. (9) of Ref. \cite{gh1},  Eq. (6)
of Ref. \cite{gh2}, and Eq. (5) of Ref. \cite{gh3}). For $\theta = 0$ he
used inappropriate
equations and for $\theta = \pi/4$ there should not be any mixing.

It is interesting to recall that using the orthogonality relation 
(\ref{con}) on the continuum  Musaelian et al \cite{mu} derived the BCS
equations for the mixed-symmetry states $s+d$ and $s+id$.   
In agreement with the present comment  and in contradiction with the
investigation by Ghosh \cite{gh1,gh2,gh3}
they (a) confirmed the existence of
mixed-symmetry states for the mixing angles $\theta =0$ and $\pi/2$ only
and 
(b) reported the BCS equation for the $s+d$  state, which is
structurally quite similar
to Eqs. (\ref{a2}) and (\ref{b2}) above. 
As the study of Musaelian et al
\cite{mu} referred to the continuum, in that work  
the discrete sum over ${\bf q}$ was replaced by the
integral over $\phi$ and the functions $\eta$ replaced by the circular
harmonics $\zeta$.

Recently, we used the correct equations (\ref{a2})  and (\ref{b2})
for a description of the 
$d_{x^2-y^2}+d_{xy}$ symmetry case \cite{ad2}, which corresponds to
 $\theta =0$  above.
The qualitative feature of the temperature dependence  of the $\Delta$'s
in that study is quite distinct from the erroneous results obtained by
Ghosh \cite{gh1,gh2} by using the inappropriate equations.  

The author thanks Dr. Manuel de Llano, Dr. Mauricio Fortes, and Dr. Miguel
Angel Sol\'is for very kind hospitality at the Universidad Nacional
Aut\'{o}noma de
M\'{e}xico. Partial support from UNAM-DGAPA-PAPIIT and
CONACyT
(M\'{e}xico), and CNPq and FAPESP\ (Brazil)
is
gratefully acknowledged.

\end{document}